\documentclass[apj]{emulateapj}
\bibpunct[; ]{(}{)}{;}{a}{}{;}
\usepackage{apjfonts}

\usepackage{graphicx}
\usepackage{natbib}

\newcommand{\Msolar}{M${_\odot}$\,}

\shorttitle{Evolution of the stellar mass function in Galactic clusters}
\shortauthors{De Marchi, Paresce \& Portegies Zwart}

\submitted{The Astrophysical Journal; Received 9 November 2009; Accepted 24 May 2010}

\begin{document}

\title{On the temporal evolution of the stellar mass function in
Galactic clusters}


\author{
Guido De Marchi,\altaffilmark{1}
Francesco Paresce,\altaffilmark{2} and
Simon Portegies Zwart\altaffilmark{3}
}

\altaffiltext{1}{European Space Agency, Space Science Department, Keplerlaan
1, 2200 AG Noordwijk, Netherlands; gdemarchi@rssd.esa.int}

\altaffiltext{2}{Istituto di Astrofisica Spaziale e Fisica Cosmica, Via 
Gobetti 101, 40129 Bologna, Italy; paresce@iasfbo.inaf.it}
 
\altaffiltext{3}{Sterrewacht Leiden, Leiden University, Postbus 9513, 2300 
RA Leiden, Netherlands; spz@strw.leidenuniv.nl}

\begin{abstract}  

We show that we can obtain a good fit to the present day stellar mass
functions (MFs) of a large sample of young and old Galactic clusters in
the range $0.1 - 10$\,\Msolar with a tapered power law distribution
function with an exponential truncation of the form $dN/dm \propto
m^\alpha \, [1 - e^{-(m/m_c)^\beta}]$. The average value of the power-law
index $\alpha$ is $\sim -2$, that of $\beta$ is $\sim 2.5$, whereas the
characteristic mass $m_c$ is in the range $0.1 - 0.8$\,M$_\odot$ and
does not seem to vary in any systematic way with the present cluster
parameters such as metal abundance, total cluster mass or central
concentration. However, $m_c$ shows a remarkable correlation with the
dynamical age of the cluster, namely $m_c/M_\odot \simeq 0.15 + 0.5
\times \tau_{\rm dyn}^{3/4}$, where $\tau_{\rm dyn}$ is the dynamical
age taken as the ratio of cluster age and dissolution time. { The
small scatter seen around this correlation is consistent with the
uncertainties on the  estimated value of $\tau_{\rm dyn}$.} We attribute
the observed trend to the onset of mass segregation via two-body
relaxation in a tidal environment, causing the preferential loss of
low-mass stars from the cluster and hence a drift of the characteristic
mass $m_c$ towards higher values. { If dynamical evolution is indeed
at the origin of the observed trend, it would seem plausible that
high-concentration globular clusters, now with median $m_c \simeq
0.33$\,M$_\odot$, were born with a stellar MF very similar to that
measured today in the youngest Galactic clusters and with a value of
$m_c \simeq 0.15$\,M$_\odot$. This hypothesis is consistent with 
the absence of a turn-over in the MF of the Galactic bulge down to the 
observational limit at $\sim 0.2$\,M$_\odot$ and, if correct, it would
carry the implication that the characteristic mass is not set by the
thermal Jeans mass of the cloud.} 

\end{abstract}


\section{Introduction}

There is general consensus that stars do not form in isolation but
rather from the fragmentation of molecular clouds that leads to star
clusters (e.g. Lada \& Lada 2003; Elmegreen 2010). This makes stellar
clusters ideal places to study the properties of star formation and its
outcome, the stellar initial mass function (IMF). However, it is also
equally well established today that a large majority of stars, in our
Galaxy and elsewhere, are not in clusters but in the field, since
clusters disrupt over time (see e.g. Gieles 2010). Therefore, any
attempt to set constraints on the star formation mechanisms from the
analysis of the present day stellar mass function (MF) of but the
youngest clusters cannot ignore the consequences of their dynamical
evolution.

This issue becomes particularly important if we want to compare the
results of star formation in physically different environments and with
various ages. The obvious example is addressing differences in the way
low-mass stars ($< 1$\,M$_\odot$) formed in globular clusters (GCs), at
$z \simeq 5$ or about 12\,Gyr ago, and in young clusters (YCs) in
the local universe. While in both cases the raw data show a broad
plateau in the mass distribution, suggesting a characteristic mass of
order a few tenths of M$_\odot$ (e.g. Elmegreen et al. 2008), until the
effects of dynamical evolution are properly understood, no 
meaningful conclusion can be drawn either on the uniformity of the  star
formation process over time or on the possible role of the  environment.
In order to cover a wide range of metallicities and initial densities
one is forced to compare stellar systems that formed a Hubble time
apart, such as the YCs and GCs mentioned above. Therefore, the
universality of  the IMF over time and  location remains as yet an
unresolved issue.

On the other hand, even though the two-body relaxation process that
governs the cluster's dynamical evolution and that leads to preferential
loss of low-mass stars is today rather well understood (e.g. Spitzer
1987), it is in general not possible to roll back the effects of
dynamics and derive the IMF from the present day MF, since we  cannot
trace back the trajectories of stars that have escaped the cluster. It
is however possible and statistically meaningful to study the evolution
of the stellar MF on a global scale by looking at the differences
between clusters at different evolutionary stages. This requires a
homogeneous sample of high quality observations of Galactic (open and
globular) clusters, treated in a uniform way and with reliable errors.
Early in this decade, this type of homogeneous study became possible for
GCs (Paresce \& De Marchi 2000), mostly thanks to the Hubble Space
Telescope. In the meanwhile, high quality data have become available for
YCs as well, mostly from wide field ground based surveys, thereby making
this study possible on a global scale.

\begin{center}
\begin{deluxetable*}{clllllrllcclc}[t]
\tablecaption{The sample of stellar clusters used in this study
\label{tab1}}
\tablehead{
\colhead{ID} & \colhead{Name} & \colhead{$\alpha$} & \colhead{$\beta$} &
\colhead{$m_c$} & \colhead{$m$ range} & \colhead{$[Fe/H]$} & 
\colhead{$\log M_{\rm tot}$} & 
\colhead{$c$} & \colhead{$\log t$} & \colhead{$\log t_{\rm dis}$} & 
\colhead{$\tau_{\rm dyn}$} & \colhead{Ref} \\
\colhead{} & \colhead{} & \colhead{} & \colhead{} &
\colhead{[M$_\odot$]} & \colhead{[M$_\odot$]} & \colhead{} & 
\colhead{[M$_\odot$]} & \colhead{} & \colhead{[yr]} & \colhead{[yr]} &
\colhead{} & \colhead{} }
\startdata
1 & $\rho$\,Oph   & $-1.9 \pm 0.2$ & $1.9 \pm 0.2$ & $0.17 \pm 0.03$ 
  & $0.02$ -- $7$ & $0.08$ & $3.3$ &   & $ 5.7$ & $8.7$&$0.001$ & a \\
2 & Orion N. C.   & $-1.9 \pm 0.2$ & $2.0 \pm 0.3$ & $0.19 \pm 0.05$ 
  & $0.02$ -- $50$ & $-0.01$ & $1.8$ & $0.3$ & $ 6.0$ & $8.7$ &$0.002$ & b, c \\
3 & Taurus        & $-2.3 \pm 0.2$ & $1.8 \pm 0.2$ & $0.51 \pm 0.05$ 
  & $0.03$ -- $3.5$& $-0.03$ & $1.7$  &       & $ 6.0$ & $6.3$ & $0.5$ & d \\
4 & IC\,348       & $-1.9 \pm 0.1$ & $3.3 \pm 0.2$ & $0.12 \pm 0.05$ 
  & $0.03$ -- $2.2$& $-0.01$ & $1.8$ & $0.3$ & $ 6.3$ & $8.7$ & $0.004$ & e \\
5 & $\sigma$\,Ori & $-2.0 \pm 0.1$ & $3.2 \pm 0.2$ & $0.14 \pm 0.02$ 
  & $0.02$ -- $2.4$& $-0.01$ & $2.3$ & $0.6$ & $ 6.5$ & $8.7$ & $0.006$ & f \\
6 & $\lambda$\,Ori& $-1.9 \pm 0.1$ & $2.3 \pm 0.2$ & $0.12 \pm 0.02$ 
  & $0.04$ -- $2.7$& $-0.01$ & $2.8$ & $0.6$ & $ 6.7$ & $8.7$ & $0.01$ & g \\
7 & Cha\,I & $-1.9 \pm 0.2$ & $2.3 \pm 0.3$ & $0.15 \pm 0.03$ 
  & $0.01$ -- $3.5$& $-0.11$ & $2.6$ &       & $ 6.7$ & $8.7$ &$0.01$ & h \\
8 & IC\,2391   & $-2.1 \pm 0.1$ & $2.4 \pm 0.3$ & $0.14 \pm 0.02$ 
  & $0.05$ -- $0.5$& $-0.09$ & $2.2$ & $0.6$ & $ 7.7$ & $8.8$ & $0.06$ & i \\
9 & Blanco\,1  & $-1.7 \pm 0.1$ & $1.8 \pm 0.2$ & $0.22 \pm 0.03$ 
  & $0.05$ -- $1.75$& $0.14$ & $3.5$ & $0.6$ & $ 8.1$ & $9.1$ & $0.09$ & j \\
10 & Pleiades  & $-2.2 \pm 0.1$ & $2.3 \pm 0.2$ & $0.27 \pm 0.03$ 
  & $0.04$ -- $9$& $0.03$ & $2.9$ & $0.6$ & $ 8.1$ & $9.3$ & $0.06$ & k \\
11 & M\,35     & $-1.7 \pm 0.1$ & $2.4 \pm 0.2$ & $0.30 \pm 0.03$ 
  & $0.09$ -- $1.4$& $-0.21$ & $3.2$ & $0.7$ & $ 8.2$ & $9.4$ & $0.08$ & i \\
12 & Coma Ber  & $-1.3 \pm 0.4$ & $1.5 \pm 0.3$ & $0.45 \pm 0.10$ 
  & $0.12$ -- $1.1$& $-0.08$ & $2.4$ & $0.4$ & $ 8.6$ & $8.9$ & $0.48$ & b \\
13 & Praesepe  & $-2.0 \pm 0.2$ & $4.0 \pm 0.5$ & $0.20 \pm 0.02$ 
  & $0.07$ -- $1.5$& $0.00$ & $3.3$ & $0.7$ & $ 8.8$ & $9.5$ & $0.18$ & l, m\\
14 & Hyades    & $-2.1 \pm 0.1$ & $2.8 \pm 0.2$ & $0.45 \pm 0.05$ 
  & $0.07$ -- $2.7$& $0.14$ & $2.6$ & $0.4$ & $ 8.8$ & $9.2$ & $0.37$ & n \\[0.15cm]
15 & NGC\,104  & $-2.0$ & $1.8 \pm 0.15$ & $0.33 \pm 0.02$ 
  & $0.11$ -- $0.7$ & $-0.76$ & $5.84$ & $2.03$ & $10.11$ & $10.93$& $0.15$ & p \\
16 & NGC\,5139 & $-2.0$ & $2.3 \pm 0.2$  & $0.33 \pm 0.02$ 
  & $0.13$ -- $0.7$ & $-1.62$ & $6.17$ & $1.61$ & $10.06$ & $10.75$ & $0.21$ & p \\
17 & NGC\,5272 & $-2.0$ & $2.4 \pm 0.2$  & $0.33 \pm 0.02$ 
  & $0.16$ -- $0.7$ & $-1.57$ & $5.65$ & $1.84$ & $10.06$ & $10.92$ & $0.14$ & p \\
18 & NGC\,6121 & $-2.0$ & $2.8 \pm 0.25$ & $0.35 \pm 0.02$ 
  & $0.09$ -- $0.5$ & $-1.20$ & $4.83$ & $1.59$ & $10.10$ & $10.16$ & $0.88$ & p \\
19 & NGC\,6254 & $-2.0$ & $1.9 \pm 0.2$ & $0.33 \pm 0.02$ 
  & $0.13$ -- $0.6$ & $-1.52$ & $5.00$ & $1.40$ & $10.06$ & $10.39$ & $0.47$ & p \\
20 & NGC\,6341 & $-2.0$ & $2.0 \pm 0.2$ & $0.30 \pm 0.02$ 
  & $0.11$ -- $0.7$ & $-2.28$ & $5.30$ & $1.81$ & $10.12$ & $10.40$ & $0.52$ & p \\
21 & NGC\,6397 & $-2.0$ & $2.3 \pm 0.2$ & $0.33 \pm 0.02$ 
  & $0.09$ -- $0.55$ & $-1.95$ & $4.65$ & $2.50$ & $10.10$ & $10.27$ & $0.67$ & p \\
22 & NGC\,6656 & $-2.0$ & $2.3 \pm 0.3$ & $0.31 \pm 0.03$ 
  & $0.10$ -- $0.6$ & $-1.64$ & $5.46$ & $1.31$ & $10.10$ & $10.62$ & $0.31$ & p \\
23 & NGC\,6752 & $-2.0$ & $2.6 \pm 0.1$ & $0.34 \pm 0.02$ 
  & $0.10$ -- $0.6$ & $-1.56$ & $5.15$ & $2.50$ & $10.07$ & $10.50$ & $0.37$ & p \\
24 & NGC\,6809 & $-2.0$ & $2.3 \pm 0.3$ & $0.30 \pm 0.02$ 
  & $0.11$ -- $0.6$ & $-1.81$ & $5.04$ & $0.76$ & $10.09$ & $10.32$ & $0.58$ & p \\
25 & NGC\,7078 & $-2.0$ & $2.7 \pm 0.3$ & $0.23 \pm 0.02$ 
  & $0.15$ -- $0.7$ & $-2.26$ & $5.75$ & $2.50$ & $10.07$ & $10.93$ & $0.14$ & p \\
26 & NGC\,7099 & $-2.0$ & $1.6 \pm 0.2$ & $0.25 \pm 0.02$ 
  & $0.15$ -- $0.7$ & $-2.12$ & $5.00$ & $2.50$ & $10.11$ & $10.39$ & $0.53$ & p \\[0.15cm]
27 & NGC\,2298 & $-2.0$ & $3.0 \pm 0.2$ & $0.54 \pm 0.01$  
  & $0.20$ -- $0.8$  & $-1.85$ & $4.50$ & $1.28$ & $10.02$ & $10.25$ & $0.59$ & q \\
28 & NGC\,6218 & $-2.0$ & $3.2 \pm 0.4$ & $0.62 \pm 0.03$  
  & $0.30$ -- $0.8$  & $-1.48$ & $4.94$ & $1.39$ & $10.10$ & $10.33$ & $0.60$ & q \\
29 & NGC\,6712 & $-2.0$ & $3.2 \pm 0.2$ & $0.80 \pm 0.03$  
  & $0.30$ -- $0.8$  & $-1.01$ & $4.97$ & $0.90$ & $10.02$ & $10.22$ & $0.63$ & q \\
30 & NGC\,6838 & $-2.0$ & $2.6 \pm 0.2$ & $0.60 \pm 0.02$ 
  & $0.30$ -- $0.8$  & $-0.73$ & $4.23$ & $1.15$ & $10.14$ & $10.21$ & $0.84$ & q\\
\enddata
\tablecomments{Table columns are as follows: ID number; cluster name;
best fitting value of $\alpha$ and $1\,\sigma$ uncertainty; best fitting
value of $\beta$ and $1\,\sigma$ uncertainty; best fitting value of
$m_c$ and $1\,\sigma$ uncertainty; mass range over which the TPL fit is
performed; cluster metallicity; total cluster mass; central
concentration, or the  logarithmic ratio of tidal radius and core
radius; cluster age; estimated time to dissolution; dynamical time, or
the ratio of the cluster age and  the time to dissolution; bibliographic
reference for the MF data, as follows. For YCs:  (a) \cite{luh99}; (b)
\cite{kra07}; (c) \cite{sle04}; (d) \cite{sal10}; (e) \cite{mue03}; (f)
\cite{cab08};  (g) \cite{bar04a}; (h) \cite{luh07}; (i) \cite{bar04b};
(j) \cite{mor07}; (k) \cite{mor03}; (l) \cite{cha05}; (m) \cite{wil95}; 
(n) \cite{bou08}. For dense GCs: (p) \cite{pdm00}. For loose GCs: (q)
\cite{dem07}. For YCs, the metallicity and age are from the bibliography
in column Ref. and references therein, whereas total mass and
concentration are from Piskunov et al. (2008), { although the value of
M$_{\rm tot}$ is from K\"uc\"uk \& Akkaya (2010) for Taurus, from Hodapp
et al. (2009) for $\sigma$\,Ori and from Dolan \& Mathieu  (2002) for
$\lambda$\,Ori.} For GCs, the metallicity, total mass and concentration
are from the 2003 revision of the GC catalogue of Harris (1996), whereas
the age is from Forbes \& Bridges (2010). The values of $t_{\rm dis}$
have been determined as explained in Section\,5, whereas $\tau_{\rm
dyn}$ is the dynamical age of each cluster, expressed as $t/t_{\rm
dis}$.}
\end{deluxetable*}
\end{center}

\section{The tapered power-law}

Since our goal is to detect and quantify changes in the shape of the MF in
different environments, in particular in GCs and YCs, we need a functional
form for the distribution of stellar masses that is flexible enough to adapt
to the variety of observed MFs and yet simple enough to be described by a
small number of parameters over a wide mass range. Kroupa (2002) has
proposed a segmented, multi-part power-law for this purpose, but we cannot
adopt it here because this approach fixes the mass points between which the
slope is fitted, thereby making it impractical or even impossible to detect
variations in the characteristic mass, which, as we will see, are the
signature of dynamical evolution.

Our choice falls instead on a tapered power-law (TPL) distribution of the
type
\vspace*{-0.1cm}
\begin{equation}
f \left(m \right)=\frac{d\,N}{d\,m}\propto m^{\alpha} \left[ 1-
e^{-(m/m_c)^\beta} \right]
\end{equation}

\noindent 
where $m_c$ is the characteristic mass, $\alpha$ the index of the
power-law portion for high masses and $\beta$ the tapering exponent that
defines the shape of the MF below the characteristic mass $m_c$. The TPL
combines in one expression the ubiquitous power-law shape above
1\,M$_\odot$ (e.g. Salpeter 1955) with the plateau and drop observed
near the H-burning limit (e.g. Paresce \& De Marchi 2000; Chabrier
2003). As we already showed (De Marchi, Paresce \& Portegies Zwart
2005), the TPL fits remarkably well the MFs of both YCs and GCs over the
entire stellar mass range.  


\begin{figure}
\centering
\hspace*{-0.8cm}\includegraphics[width=3.8in]{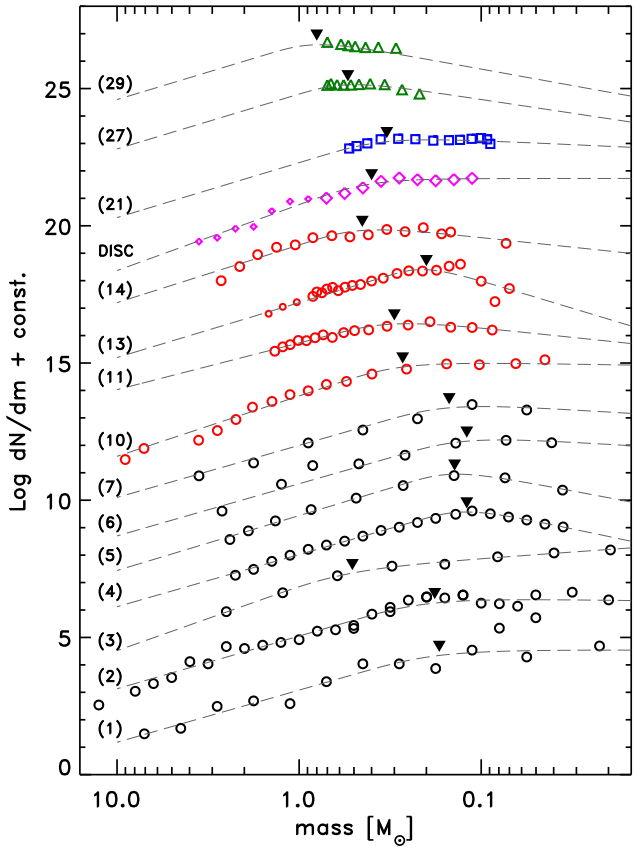}
\caption{Mass functions of selected clusters in our sample (see
Table\,\ref{tab1} for the ID numbers). The data for YCs are shown as
large circles (except for Praesepe, where also small circles are used to
show the MF at the high-mass end as measured by Williams et al. (1995); 
NGC\,6397 is shown as representative for the class of dense GCs 
(squares); while the MFs of loose GCs are indicated by triangles (only
NGC\,2298 and NGC\,6712 are shown). The typical $1\,\sigma$ uncertainty
on the MF is $0.5$\,dex at $\sim 0.3$\,\Msolar. The  dashed lines
represent the best fitting TPL distributions, as per the  parameters of
Table\,\ref{tab1}, with the values of $m_c$ indicated by the dark
down-pointing triangles. For reference, we also show the MF of the
Galactic disc. At the low-mass end the data are those recently obtained
by  Covey et al. (2008) from Sloan Digital Sky Survey and Two Micron All
Sky Survey data (large diamonds), whereas above $\sim 1$\,\Msolar we
show the MF obtained by Reid et al. (2002) and corrected for the effects
of stellar evolution (small diamonds). }
\label{fig1}
\end{figure}

\vspace*{1cm}
\section{The data sample}

The data used in this work come from a number of different sources. As
regards GCs, we considered the entire HST sample of Paresce \& De Marchi
(2000), which is made up of twelve relatively dense clusters, and added to
it four low-concentration GCs studied by our team in recent years
(for details see Table\,\ref{tab1}). De Marchi et al. (2007) showed that a
strong trend exists between a cluster's central concentration $c$
(defined as the logarithmic ratio of the tidal radius and core radius)
and the shape of its present stellar MF: dense GCs, i.e. those with $c>1.5$,
have a relatively steep MF with an average power-law index of $-1.5$ in
the range $0.3 - 0.8$\,\Msolar, while loose GCs with $c<1.5$ show a
much flatter or even inverted MF over the same mass range. Several
interpretations exist for this result (e.g. Baumgardt, De Marchi \&
Kroupa 2008; Kruijssen \& Portegies Zwart 2009; Vesperini, McMillan \&
Portegies Zwart 2009; Kruijssen 2009), but they all point to a shorter
time to dissolution for loose GCs. For this reason, it is important to
cover both dense ($c > 1.5$) and loose ($c< 1.5$) clusters when studying
the temporal evolution of the stellar MF. 

As regards YCs, the data cover high quality observations of fourteen
star-forming regions and associations of various ages, as  shown in
Table\,\ref{tab1}. For each YC, we searched for the most accurate
determination of the MF in the recent literature (see references in
Table\,\ref{tab1}), looking specifically for information on the number
of stars measured in each mass bin. In those cases in which  the MFs
were not directly available in tabular form, we extracted the data
points (i.e. number of stars in each mass bin) from the published
graphs, but we specifically ignored any fits (with power-law or
log-normal functions) or interpolations of these data that the various
authors may have carried out. The MFs in our sample were determined
either by converting an observed luminosity function via a theoretical
mass--luminosity relationship or by counting the number of stars falling
between evolutionary tracks in a color--magnitude or
Hertzsprung--Russell diagram. We only  restricted our search to works in
which the MF is based on known cluster members and a correction for
photometric incompleteness has been applied to the luminosity function,
if needed. Furthermore, to guarantee that the sample is as homogeneous
as possible, we have ignored  any corrections to the MF meant to account
for the presence of binaries, since these are rather uncertain and only
available in a limited number of cases. Therefore, the MFs that we
consider in this study, for both YCs and GCs, are the MFs of stellar
systems, i.e. we do not distinguish between single  and multiple stars.

In order to prevent biases in the results, the data should refer to the
global MF (GMF), i.e. to the MF of the cluster as a whole. For GCs,
where complete cluster coverage is not always possible, we used
information on mass stratification and mass segregation to derive the
GMF from the local MF (see e.g. De Marchi et al. 2006). Alternatively,
we used the MF measured near the half-light radius (i.e. the
effective radius containing half of  the cluster's luminosity, see e.g.
Portegies Zwart, McMillan \& Gieles 2010), since it has been shown to
reflect quite reliably the properties of the GMF (De Marchi et al.
2000). For YCs, we specifically selected those objects and studies for
which the coverage is as complete as possible.

We then performed a multivariate fit to each MF with a TPL
distribution and derived the values of the parameters $\alpha$, $\beta$
and $m_c$ that simultaneously provide the smallest residuals. The values
of the best fitting parameters and their associated $1\,\sigma$
uncertainties are given in Table\,\ref{tab1} for all the clusters in our
sample. The MFs of a few selected clusters are also shown graphically in
Figure\,\ref{fig1}, together with the best TPL fits as per the
parameters of Table\,\ref{tab1}. We compare and discuss the results of 
our best fits in Section\,5, but we precede that with a short discussion 
on the value of $\alpha$.

\section{Mass function shape above $\sim 0.8$\,\Msolar}
 
In the TPL distribution described by Equation\,1, the value of
$\alpha$ determines the shape of the MF above the characteristic mass
$m_c$. Since the data for GCs do not constrain the mass range above the
main-sequence turn-off ($> 0.8$\,M$_\odot$), we had to assume a value of
$\alpha$ for those objects. An obvious first choice would be the slope
of the  Salpeter (1955) IMF, namely $\alpha=-2.35$. According to Kroupa
(2002), that is the average value of the slope of the stellar MF in
clusters and associations above $0.5$\,\Msolar. However, this is not a
meaningful  average because of the very large scatter at $ m>
1$\,\Msolar displayed by the MF slopes included in Kroupa's (2002)
compilation, which in turn are taken from  Scalo (1998). In fact, after
discussing with this latter author (J. Scalo, priv. comm.), we realised
that the $\alpha=-2.35$ value obtained in that way is not a meaningful
average. It is also not consistent with any of the individual $\alpha$
values that we measure for the YCs in Table\,\ref{tab1}, except for
Taurus and possibly Pleiades.

As an alternative we could use the MF of the Galactic disc, but decided
against it because of the large uncertainties involved. The most recent
determination is that by Covey et al. (2008) using Sloan Digital Sky
Survey and 2 MASS data, although these observations do not reach above
$\sim 1$\,\Msolar (see large diamonds in Figure\,\ref{fig1}). To extend
the MF to higher masses, these authors use the measurements of Reid,
Gizis \& Hawley (2002). We do the same and show with small diamonds in
Figure\,\ref{fig1} the MF that Reid et al. (2002) obtained using an
empirical mass--magnitude relationship and after correction for the
effects of stellar evolution (this provides them with an estimate of 
the IMF). The best fit with a TPL distribution gives $\alpha = -2.4 \pm
0.4$, $\beta = 2.3 \pm 0.4$ and $m_c=0.4 \pm 0.05$, revealing again a
rather large uncertainty around the value of $\alpha$.

In the end, following the suggestion of an anonymous referee, we
concluded that the most appropriate value of $\alpha$ to adopt for GCs
is the average of the best fitting $\alpha$ values that we obtain in the
same mass range for YCs, namely $\alpha= -1.97 \pm 0.17$. This
is also  conveniently very close to the value that Van Buren (1985)
found for the MF slope in the range $5-120$\,\Msolar from a detailed
analysis of the all-sky sample of O stars of Garmany, Conti \& Chiosi
(1982) that makes use of high-resolution extinction maps derived
spectroscopically to account for the effects of obscuration on star
counts (Scalo 1986).  

Finally, we note that observational constraints on the initial value of
$\alpha$ for GCs could in principle come from the analysis of the white
dwarf luminosity function. The results of Richer et al. (2004) on the
white dwarf cooling sequence of M\,4 (NGC\,6121) seem to suggest a value
in the range $-2.5 < \alpha < -1.5$, but more recent work by the same
team (Richer et al. 2008) on another cluster, NGC\,6397, seems to
require a much steeper slope, possibly as large as $\alpha \simeq -5$.
Therefore, it appears that the uncertainties are presently still too
large for this method to place meaningful empirical constraints on the
value of $\alpha$ for GCs, although progress in this field is likely
with the advent of the James Webb Space Telescope in the near future.

\section{Evolution of the mass function}
 
The mean $\alpha$ and $\beta$ values for the YCs in our sample\footnote{
We ignore Coma Ber when calculating the average of $\alpha$
and $\beta$ for YCs because the limited number of independent data
points in its MF results in rather large uncertainties on the fitting
parameters.}, resulting from our best fits shown in Table\,\ref{tab1},
are respectively $-1.97 \pm 0.17$ and $2.51 \pm 0.62$ ($1\,\sigma$
standard deviation uncertainties). For GCs, having assumed $\alpha
\equiv -2$, we find $\beta=2.49 \pm 0.47$. In spite of the widely
different physical properties of the  clusters in our sample, e.g. in
terms of chemical composition, total mass, stellar density and age (see
Table\,\ref{tab1}), the values of $\alpha$ and $\beta$ span a relatively
narrow range and, in particular, there is no significant difference in
$\beta$ between YCs and GCs. Obviously, having assumed $\alpha \equiv
-2$ for GCs has in principle a systematic effect on $\beta$, but this
does not seem to be large (an even smaller effect on $m_c$ is discussed
below). For example, with our original assumption of $\alpha \equiv
-2.35$ we found $\beta = 2.6 \pm 0.3$. This suggests that that there
does not seem to be substantial difference in the way the stellar MF 
rises to and falls from its peak, i.e. the characteristic mass, in 
clusters covering a wide range of physical properties, including age,
chemical composition, total mass and density. 

Where large differences are seen, however, is in the value of the 
characteristic mass itself. The average value of $m_c$ for dense GCs is
$0.31 \pm 0.03$\,M$_\odot$ and for loose globulars it grows to $0.64 \pm
0.10$\,M$_\odot$, although the statistics here is limited. As noted
above, the actual value of $m_c$ depends in principle on the assumed
$\alpha$, but changing the latter by $0.4$ would cause  $m_c$ to vary by
just $0.01$\,\Msolar for the GCs in our sample. It is therefore unlikely
that the differences in the $m_c$ values that we observe are just the
result of uncertainties in $\alpha$. The variation in $m_c$ is equally
large for YCs, ranging from $0.12$\,M$_\odot$ to $0.51$\,M$_\odot$, and
here the data provide much more solid constraints on the value of
$\alpha$.

In principle, all these $m_c$ variations could reflect differences in
the star formation process, in turn resulting from differences in the
physical conditions of the environment (e.g. Larson 2005 and references
therein). However, as the data in Table\,\ref{tab1} show, there does not
seem to be any correlation between the value of $m_c$ and the clusters'
physical properties such as total mass, concentration, age or
metallicity. 

For example, the dense GCs in our sample span almost two decades in
metallicity and total mass, but have practically the same $m_c$, while
all YCs have practically the same metallicity but a wide range of $m_c$
values. Therefore, if the environment plays a role in determining the 
shape of the stellar MF this effect { does not seem to} be the
dominant one, at least not for the clusters in our sample. In fact,
taken at face value, the data in Table\,\ref{tab1} show no sign of such
an effect.

A more interesting possibility to consider in order to explain the
observed spread in $m_c$ is the role of dynamical evolution, which could
alter the shape of the MF over time because of the preferential loss of
low-mass stars caused by the two-body relaxation process. It is thus
worth investigating whether our data show any correlation or trend
between the characteristic mass and the dynamical state of the cluster.

For this reason we have listed in Table\,\ref{tab1} the dynamical age
$\tau_{\rm dyn}$ of each cluster, expressed as $t/t_{\rm dis}$ or the
cluster age $t$ in units of the dissolution time $t_{\rm dis}$, in turn
defined as the time at which the cluster has lost 95\,\% of its original
mass. Using N-body models, Baumgardt \& Makino (2003; hereafter BM03)
showed that $t_{\rm dis} \propto t^x_{\rm rh} t^{1-x}_{\rm cr}$ where
$t_{\rm rh}$ is the half-mass relaxation time, $t_{\rm cr}$ the crossing
time and $x$ depends on the initial concentration of the cluster $W_0$.
Later, Gieles et al. (2005), using the same N-body models, found that
$t_{\rm dis}$ can also be expressed as a function of the total initial
mass of the cluster $M_{\rm ini}$ as $t_{\rm dis} \propto t_0 M_{\rm
ini}^{0.62}$ (see also Gieles 2010), where the tidal disruption
parameter $t_0$ depends on the properties of the host galaxy and on the
cluster's orbit. 

The values of $\tau_{\rm dyn}$ listed in Table\,\ref{tab1} come from
Baumgardt et al. (2008) for the specific GCs in our sample {and take
into account the differences that exist between their orbital
parameters. As regards the YCs in the sample, they are all located in
the solar neighbourhood and should therefore experience rather similar
interactions with the Galaxy. This is consistent with the results of
Lamers et al.'s (2005) study of the age distribution of open clusters in
the solar neighbourhood within 600\,pc, based on the catalogue by
Kharchenko et al. (2005). They conclude that $t_0 =
3.3^{+1.5}_{-1.0}$\,Myr, so if we accept an uncertainty of $\sim
${\small $1/3$} on the value of $t_{\rm dis}$, we can ignore its
dependence on the specific cluster's orbit. The individual $t_{\rm dis}$
values for the YCs in our sample can then be derived following BM03's
N-body simulations for clusters with low initial concentration $W_0=5$
and a distance of $R_G=8.5$\,kpc from the Galactic centre.

Using BM03's Figure\,1a, we have estimated the initial mass $M_{\rm
ini}$ of our clusters from their present-day total mass $M_{\rm tot}$
and age  $t$ as given in Table\,\ref{tab1}. While BM03 consider clusters
with an initial number of stars $N$ ranging from 8K to 128K objects, for
ages younger than $\sim 650$\,Myr, as is the case for all YCs in our
sample, the ratio $M_{\rm tot}/M_{\rm ini}$ varies only marginally with
$N$, at most by 10\,\% and only imperceptibly for $t < 300$\,Myr. We
have therefore followed the curve for $N=$32K in Figure\,1a of BM03 to 
derive the estimated value of $M_{\rm ini}$, which we have then used to
derive the time to dissolution $t_{\rm dis}$ and hence $\tau_{\rm dyn}$
by means of Figure\,3 in that same paper, again following the
distribution for clusters with $R_G=8.5$\,kpc and $W_0=5$. For clusters
younger than 100\,Myr we have assumed  $t_{\rm dis} \simeq 500$\,Myr but
our conclusions would not change if $t_{\rm dis}$ were larger or smaller
by a factor of two for these clusters.

It is to be expected that the values of $\tau_{\rm dyn}$ derived in this
way are somewhat uncertain, even though they are as accurate as one can
obtain based on the available information. Lamers et al. (2005) point
out that  the disruption times obtained by BM03 are most likely an
upper limit  because they are calculated for tidal disruption in a
smooth tidal field without other destruction mechanisms. Hereafter we
will assume an uncertainty of about $45\,\%$ on $t_{\rm dis}$, which
combined with a typical accuracy of $\sim 20\,\%$ on cluster ages
implies an uncertainty of about $50\,\%$ on $\tau_{\rm dyn}$.} 

The relatively high $\tau_{\rm dyn}$ value of the Hyades is 
consistent with the conclusion of Perryman et al. (1998) that this
cluster has lost about {\small $3/4$} of its original mass. Conversely, 
Praesepe
might not be as dynamically old as the estimated  value of $\tau_{\rm
dyn}$ would suggest if it is confirmed that the cluster is indeed the
result of a merger of two clusters with different ages (see Holland et
al. 2000; Franciosini, Randich \& Pallavicini 2003). Finally, a
particularly interesting case is that of Taurus. Ballesteros--Paredes et
al. (2009) conducted a gravitational analysis of its orbit and concluded
that the Taurus molecular cloud must be suffering significant tidal
disruption, in spite of its young age (1\,Myr). We conservatively
assigned to it $\tau_{\rm dyn} =$~{\small $1/2$}, although the actual 
value could be larger.

\begin{figure}[t]
\centering
\hspace*{-0.8cm}\includegraphics[width=3.8in]{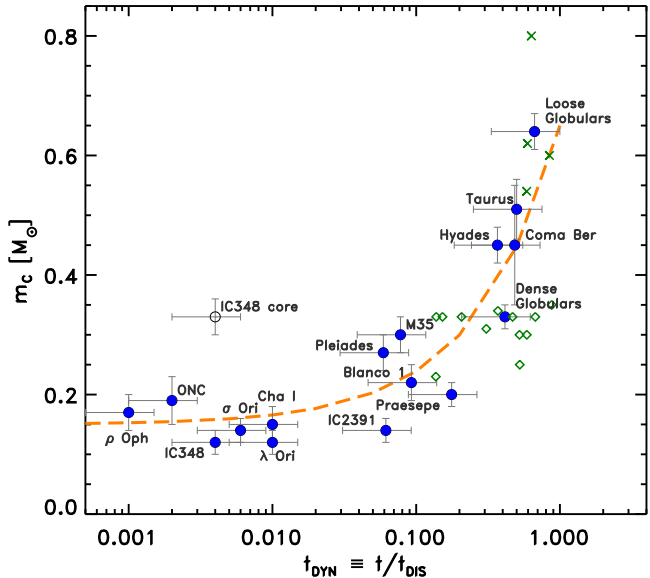}
\caption{Run of the characteristic mass $m_c$ as a function of the
dynamical age $\tau_{\rm dyn}$. Dense and loose GCs are indicated,
respectively, with diamonds and crosses, with a filled circle providing
the average value of each class. The core of IC\,348 is indicated as an
open circle to distinguish it from the whole cluster. The dashed line is
an eye-ball fit to the data.}
\label{fig2}
\end{figure}

The run of $m_c$ as a function of $\tau_{\rm dyn}$ is shown in
Figure\,\ref{fig2}, where all clusters are labeled individually, except
for the GCs (see caption for details). The figure reveals a rather
remarkable trend of increasing $m_c$ with dynamical age over the entire
range covered by the data. The dashed line is a purely empirical
eye-ball fit to the data with the relationship

\begin{equation}
m_c / M_\odot \simeq 0.15 + 0.5 \times \tau_{\rm dyn}^{3/4}
\end{equation}

\noindent although an index of {\small $1/2$} would still give an acceptable
fit. Equally remarkable in Figure\,\ref{fig2} is the absence of
dynamically young clusters with large $m_c$ but even more striking,
considering the logarithmic scale of the abscissa, is the lack of
dynamically old objects ($\tau_{\rm dyn} > 0.3$) with a small $m_c$,
even for GCs with a total mass of order $10^6$\,\Msolar. 

{There are several effects that can cause the scatter seen in
Figure\,\ref{fig2} around the dashed line, including the uncertainty on
$\tau_{\rm dyn}$ and $m_c$, the role of binaries and incomplete coverage
of the clusters. As an example of the latter, we show in
Figure\,\ref{fig2} also the data point corresponding to the sole central
regions of IC\,348 (from the MF of Muench et al. 2003). In spite of the
young age of the cluster, the MF in its core shows strong signs of mass
segregation, with $m_c=0.33$\,M$_\odot$ as opposed to
$m_c=0.12$\,M$_\odot$ for the cluster as a whole. This may be due to
primordial mass segregation and/or to very fast dynamical evolution (see
e.g. Allison et al. 2009), but in either case this data point should be
excluded from the sample in order to not skew the results. In spite of
our attempts to select only clusters with complete radial coverage,
however, this information might not be completely available or fully
correct in the literature. This is likely to cause a larger uncertainty
on $m_c$ than what we obtain through the TPL fit to the MF and could 
account for some of the residual scatter that we observe. An additional
source of uncertainty on $m_c$ comes from the fact that all the MFs
considered here are system MFs, i.e. we do not attempt to correct for
the presence of binaries. For dense GCs the binary fraction is just a
few percent (Davis et al. 2008) and is not expected to significantly
alter the value of $m_c$, but it is very likely higher for loose GCs and
YCs (Sollima et al. 2007; 2010) and its effect should be considered. We
will do so in a forthcoming paper. Finally, as mentioned above the
uncertainty on the value of $\tau_{\rm dyn}$ can be important and this
is witnessed by the size of the error bars that we show in
Figure\,\ref{fig2}.}

\section{Discussion and conclusions}

At least qualitatively, the trend seen in Figure\,\ref{fig2} is fully
consistent with the onset of mass segregation via two-body relaxation in
a tidal environment, causing the preferential loss of low-mass stars
from the cluster and hence a drift of the characteristic mass towards
higher values. Vesperini \& Heggie (1997) showed from N-body models that
stellar evaporation, integrated over the cluster's orbit and further
enhanced by the presence of the Galactic tidal field, causes a
flattening of the MF, i.e. a selective depletion at the low-mass end.
Their models do not reveal a drift of the characteristic mass towards
higher values, but their IMF in all cases is a pure power-law. A drift
of $m_c$ is visible in the N-body models of Portegies Zwart et al.
(2001), although it proceeds very slowly and only appears when the study
of the MF is limited to the regions inside the half-mass radius. 

{That mass segregation is present in both young and globular
clusters has long been established observationally. Examples include
many of the YCs in our sample, such as M\,35 (e.g. Mathieu 1983; Barrado
y Navascu\'es et  al. 2001), Pleiades (e.g. Converse \& Stahler 2008),
Hyades (Reid 1993; Perryman et al. 1998), Blanco\,1 (Moreaux et al.
2007). As for GCs, mass segregation has been observed in all those
objects where MF measurements exist at various radial distances (see
e.g. De Marchi et al. 2007 and references therein) and can be
satisfactorily explained as being the result of the two-body 
relaxation process (Spitzer 1987). Furthermore, De Marchi et al. (2007)
have also shown that dense and loose GCs have today systematically
different GMFs and this is consistent with a much stronger selective
loss of low-mass stars in the latter. Loose GCs have systematically
larger $m_c$ than denser clusters with otherwise similar metallicity or
space motion parameters. This strongly suggests that the value of $m_c$
must drift as a result of the cluster's dynamical evolution, which is
quicker in loose systems, even though the analysis of the N-body models
(e.g. Portegies Zwart et al. 2001) has so far failed  to detect this
effect.} In fact, Kruijssen (2009) has recently developed a simple
physical model for the evolution of the MF that builds on the pioneering
work of H\'enon (1969). His results (see Figure\,14 in Kruijssen 2009)
show a shift in the inflection point of the MF with evolutionary time,
consistent with a drift in $m_c$ (D. Kruijssen, priv. comm.).  

{ A potentially very interesting consequence of the trend shown in
Figure\,\ref{fig2}, if dynamical evolution is indeed at its origin, is
that it would seem plausible that all GCs, having now $m_c \simeq
0.33$\,M$_\odot$ if they are dense or $\sim 0.65$\,M$_\odot$ if they are
loose, were born with a stellar MF like the one measured today in the
youngest Galactic clusters, or more precisely with $m_c \simeq
0.15$\,M$_\odot$. Although not necessarily required by
Figure\,\ref{fig2}, this scenario is consistent with it.

The careful reader will notice that this statement appears at odds with
the conclusions of a previous work by our team. From a study of the GMFs
of twelve halo GCs with widely different dynamical histories and noting
that they show a very similar characteristic mass ($m_c=0.33 \pm 
0.03$\,\Msolar), Paresce \& De Marchi (2000) concluded that this value
of $m_c$ had to be a feature of the stellar IMF in GCs. However, it is
now apparent that this conclusion was the result of a selection effect,
due to the fact that all twelve GCs in the Paresce \& De Marchi (2000)
sample have high central concentration, since no reliable information
on the low-mass end of the MF of loose GCs was available at that time.
The analysis presented by De Marchi et al. (2007) suggests that dense
GCs have lost less low-mass stars than looser clusters throughout their
life, but this does not mean that they have lost none. Therefore, one
can no longer conclude that the $m_c \simeq 0.33$\,M$_\odot$ value
observed in the present-day GMF of dense GCs reflects the characteristic
mass of their IMF, since it could have been smaller. }

The possibility that GCs were born with a much smaller $m_c$ value than
what we measure today in their MF { would also be consistent} with
the absence of a turn-over in the MF of the Galactic bulge down to the
observational limit at $\sim 0.2$\,M$_\odot$ (Zoccali et al. 2000).
Regardless as to whether the bulge is primarily the result of fast
formation at early epochs (e.g. Ballero et al. 2007) or a collection of
the stars lost from disrupted clusters throughout the life of the Galaxy
(e.g. Gnedin \& Ostriker 1997), its very location at the bottom of the
Galaxy's potential well makes it very hard for low-mass stars to escape
from it. Therefore, while the present MF of the Galactic disc could
represent a complex average over time and space, that of the bulge
should have not been altered by dynamical evolution and, unlike the case
of the comparably old GCs, should still reflect the properties of the
IMF. Forthcoming HST observations of the Galactic bulge (see e.g. Brown
et al. 2009) will provide insights on its MF below $0.15$\,M$_\odot$,
where we expect a turn-over.

{To be sure, N-body models of dense stellar systems able to deal
with a large number of particles ($N > 10^6$) and incorporating a
realistic treatment of stellar evolution and of the interaction with the
Galactic tidal field (see Portegies Zwart et al. 2010) would be
necessary to simulate the growth of the characteristic mass with time in
GCs. Our analysis (see Section\,5) suggests that young and globular
clusters may share rather similar values of the $\alpha$ and $\beta$
parameters. Thus, if the picture sketched above is correct and, by
strongly affecting the value of $m_c$, dynamical evolution act as the
main source of variations in the present-day stellar MF of Galactic
clusters, one might postulate that Population I and II stars should have
similar IMFs. While not necessarily required by our analysis, this
hypothesis is fully compatible with our results. Claims of the IMF
universality that one finds in the literature (e.g. Gilmore 2001;
Elmegreen et al. 2008) are based on the observation that stellar MFs in
different environments have {\em similar} $m_c$ values, within a factor
of a few. Here we show that it is plausible that the original value of
$m_c$ be actually {\em the same} for Population I and II stars. 

If this is true, one would have to consider the possibility that the
characteristic mass in the IMF is not necessarily set by the Jeans mass
scale in the forming cloud (e.g. Larson 1998; Larson 2005) and that the
physical conditions of the environment do not play a significant role in
the outcome of the star formation process, i.e. on the IMF. This,
however, would not necessarily mean that the star formation process
itself is the same: recent numerical simulations proposed by Goodwin \& 
Kouwenhoven (2009) and Bate (2009) suggest that the IMF shape is largely
insensitive to parameters such as the binary fraction and mass ratio in
binaries. If this is the case, the role of the IMF as an effective tool
to probe the physics of star formation may have to be critically
reconsidered.}

\acknowledgements

We are indebted to John Scalo and to an anonymous referee for
illuminating comments and suggestions for improving the presentation of
our work. Part of this research was carried out while GDM participated
in the programme ``Formation and evolution of Globular Clusters'' at the
Kavli Institute for Theoretical Physics (UCSB, Santa Barbara,
California). FP  is grateful to the Space Science Department of ESA for
their hospitality via the visitor programme. SPZ acknowledges the
support of NWO through grant \# 639.073.803.

\end{document}